\documentclass[12pt]{article}

\usepackage{natbib}

\bibpunct[; ]{(}{)}{,}{a}{}{;}

\begin{document}

\begin{center} {\large {\bf Lema\^{i}tre's Hubble relationship}}
\end{center}

The detection of the expansion of the Universe is one of the most important
scientific discoveries of the 20th century.  It is still widely held that in
1929 Edwin Hubble discovered the expanding Universe \citep{Hubble1929}
and that this
discovery was based on his extended observations of redshifts in spiral nebulae.
Both statements are incorrect. There is little excuse for this, since there
exists sufficient well-supported evidence about the circumstances of the
discovery. The circumstances have been well documented even recently with the
publication of two books: \cite{Bartusiak2010,NB2009}.
Both were positively reviewed in the December 2009 issue of PHYSICS TODAY (page 51).
Other writers have stated the facts correctly as well
\citep[e.g.][]{Peebles1984}.

\cite{Friedman1922} was the first to publish non-static
solutions to Albert Einstein's field equations. However, he did not extend that into a
cosmological model built on astronomical observations. Some five years later,
Georges Lema\^{i}tre also discovered dynamical solutions \citep{Lemaitre1927}.
In the same publication in which he reported his discovery, he extracted
(on theoretical grounds) the linear relationship between velocity $v$ and
distance $r$: $v=$H$r$.
Combining redshifts published by \cite{Stromberg1925} (who relied mostly on
redshifts from Vesto Slipher \citep[e.g.][]{Slipher1917}) and Hubble's distances from
magnitudes \citep{Hubble1926}, he calculated two values for the ``Hubble constant'' H,
575 and 670 km sec$^{-1}$ Mpc$^{-1}$, depending on how the data is
grouped. Lema\^{i}tre concluded from those results that the Universe was expanding.
Two years later Hubble found the same velocity--distance relationship
on observational grounds \citep{Hubble1929} from practically the same observations that
Lema\^{i}tre had used. However, Hubble did not credit anyone for
the redshifts, most of which again came from Slipher.
 
Several of today's professional astronomers and popular authors
\citep[e.g.][]{Singh2005} believe that the entirety of
Lema\^{i}tre's 1927 French-language paper was re-published in
English \citep{Lemaitre1931} with the help of Arthur Eddington.
That is also incorrect; the two pages from the 1927 paper that contain
Lema\^{i}tre's estimates of the
Hubble constant are not in the 1931 paper, for reasons that have never been
properly explained.

Unfortunately several prominent people
writing in the popular press continue to promote Hubble's discovery of
the expansion of the Universe. See, for example, Brian Greene's Op-Ed piece
in the {\it New York Times} on 15 January 2011.

There is a great irony in these falsehoods still being promoted today. Hubble
himself never came out in favor of an expanding Universe; on the contrary,
he doubted it to the end of his days. It was Lema\^{i}tre  who
was the first to combine theoretical and observational arguments to show that
we live in an expanding Universe.

\bigskip
\bigskip

\noindent {\it Michael Way is a research scientist at the NASA Goddard Institute
for Space Studies in New York City and an Adjunct Professor at Hunter College,
City University of New York, USA}
\bigskip

\noindent {\it Harry Nussbaumer is an Astronomy professor emeritus at
ETH Zurich, Switzerland. He recently published as first author the book
``Discovering the Expanding Universe''.}

\end{document}